\documentclass[12pt]{iopart}
\usepackage{graphicx}
\begin{document}

\title[Static Cosmological Model]{A Static Cosmological Model Based on the Group of Conservative Transformations}

\author{Edward Lee Green }

\address{University of North Georgia, Dahlonega, GA 30597}
\ead{egreen@ung.edu}
\begin{abstract}
The group of Conservative transformations is an enlargement of the group of diffeomorphisms which leads to a richer geometry than that of general relativity.  The field variables of the theory are the usual orthonormal tetrads and also internal space tetrads.  Using the fundamental geometric object which is the curvature vector, an appropriate Lagrangian has been defined for both free-field and fields with sources.  Solutions to the corresponding field equations have been developed.  In this paper we use the static spherically symmetric tetrad field with sources to model the universe. Our fundamental assumption is that the total density comprised of both ordinary and dark matter should be constant.  The resulting model with one adjustable parameter predicts that ordinary matter is approximately 77\% of the total mass content, but this percentage is near 0\% for regions near the center of the universe.  The space is approximately isotropic for $r$ near zero. The radial and tangential pressures are negative and unequal.  The redshift is also modeled without the expanding universe and an explanation of the value of the cosmological constant is given.  Equations governing particle motion are also derived which can produce a repulsive effect and produce even larger redshifts.  Finally, the cosmic microwave background and its anisotropies are addressed and heuristic arguments are given that suggest that our theory is not inconsistent with these observations.  These results add further confirmation that the theory developed by Pandres is the fundamental theory of physics.
\end{abstract}

\submitto{}
\maketitle

\section{Introduction}

Our theory is based on a 4-dimensional space with orthonormal tetrad $h^{i}_{\,\,\mu}$ and internal tetrad $L^i_{I}$.  The metric on the world space is defined by $g_{\mu\nu}=\eta_{ij}\, h^i_{\,\,\mu}h^j_{\,\,\nu}$ where $\eta_{ij} =
diag\bigl\{-1,1,1,1\bigr\}$.   Let $\tilde{V}^\alpha$ be a vector density of
weight $+1$.  The following condition (1) defines the conservation group, a group of coordinate transformations that include the group of diffeomorphisms as a proper subgroup \cite{Pand81}:
\begin{equation}x^\nu_{\;\; ,\overline{\alpha}}\bigl(x^{\overline{\alpha}}_{\;\; ,\nu ,\mu} -
x^{\overline{\alpha}}_{\;\; ,\mu , \nu} \bigr)\; = \; 0 \qquad .  \label{conserv} \end{equation} This group preserves the wave equation and other conservation laws which are of the form $V^\alpha_{\,\, ;\alpha}=0\,$ \cite{Pand84}.

The geometry determined by (\ref{conserv}) is more general than a manifold \cite{Pand81,Pand84,Pand09,Green09,PG03}.
One must generally regard $x^{\bar{\mu}}$ as anholonomic coordinates if one retains the manifold view(see \cite{PG03}).  Alternatively, we may regard a conservative, non-diffeomorphic transformation as a mapping between two manifolds \cite{Green11}.

The geometrical content is determined by the curvature vector defined by
\begin{equation}  C_\alpha \equiv \; h_i^{\,\,\nu}\bigl(h^i_{\,\, \alpha ,\nu}-h^i_{\,\, \nu ,\alpha}
\bigr) \; = \; \gamma^\mu_{\;\;\alpha\mu} \label{curv} \end{equation}
where the Ricci rotation coefficient is given
by $\gamma^i_{\;\;\mu\nu}=h^i_{\;\mu ;\nu}$ \cite{Pand81,Pand84,Pand09}. The curvature vector, $C_\alpha$, is covariant under transformations from $x^\mu$ to $x^{\overline{\mu}}$ if and only if the transformation is conservative, i.e., it satisfies (\ref{conserv}).  A suitable scalar Lagrangian for the free field \cite{Green11} is given by
\begin{equation} {\cal L}_f= \frac1{16\pi}\int C^\alpha C_\alpha \, h \; d^4 x  \label{lagrang} \end{equation} where $h = \sqrt{-g}$ is the determinant of the tetrad.

In order to unify gravity with the electroweak and strong forces, an extension of the field variables \cite{Green09} may be needed.  We propose that $h^i_{\;\mu}$ and the internal vectors $\Lambda^i_I$ are the necessary field variables.  For the present discussion, we will need only the tetrad field, $h^i_{\;\mu}$.

    Using the Ricci rotation coefficients, one finds that
\begin{equation}C^\alpha C_\alpha = R + \gamma^{\alpha \beta \nu} \gamma_{\alpha \nu \beta}-2C^\alpha_{\; ;\alpha}
\quad , \label{CmuCmu}\end{equation}
where $R$ is the usual Ricci scalar curvature. (Note: our sign conventions are the same as Misner, Thorne and Wheeler \cite{MTW}.)  Thus, the Lagrangian density of the free field contains additional terms which may correspond to other forces or dark matter \cite{Pand09,Green09,Green11}.

Setting the variations of ${\cal L}_f$ with respect to $h^i_{\;\;\mu}$ equal to zero yields the field equations
\begin{equation}  C_{\mu ; \nu}-C_\alpha \gamma^{\alpha}_{\;\mu\nu}-g_{\mu\nu}C^{\alpha}_{\; ; \alpha} - \frac12 g_{\mu\nu} C^\alpha C_\alpha = 0 \quad .  \label{fieldeq} \end{equation}

An identity for  the Einstein tensor is
\begin{eqnarray}  G_{\mu\nu}= & C_{\mu ; \nu}- C_\alpha \gamma^\alpha_{\; \mu\nu} -g_{\mu\nu}C^\alpha_{\; ;\alpha}-\frac12 g_{\mu\nu}C^\alpha C_\alpha   \nonumber \\ \quad & +\gamma^{\;\;\alpha}_{\mu \;\; \nu ;\alpha}+\gamma^\alpha_{\;\;\sigma
\nu} \gamma^\sigma_{\;\; \mu \alpha} + \frac12 g_{\mu\nu}\gamma^{\alpha \beta
\sigma} \gamma_{\alpha \sigma \beta} \nonumber \end{eqnarray}  Using (\ref{fieldeq}) we see that the field equations may be also expressed in the form
\begin{equation}G_{\mu\nu} = \gamma^{\;\;\alpha}_{(\mu \;\; \nu) ;\alpha}+  \gamma^\alpha_{\;\;\sigma (\nu} \gamma^\sigma_{\;\; \mu) \alpha}  + \frac12 g_{\mu\nu}\gamma^{\alpha \beta
\sigma} \gamma_{\alpha \sigma \beta} \; \equiv \; 8\pi \bigl(T_{\rm f}\bigr)_{\mu\nu}  \end{equation}
with free field stress energy tensor $\mathbf{T}_{\rm f}$ and where the parentheses indicate a symmetrization. The terms of $\mathbf{T}_{\rm f}$ suggest that, when interpreted in Riemannian geometry \cite{MTW,Wein72,dFC90}, there are additional terms that could be due to dark matter or dark energy \cite{Green11}.

When sources are present, the Lagrangian is of the form
\begin{equation} {\cal L}= {\cal L}_{\rm f}+{\cal L}_{\rm s} =  \int \biggl( \frac1{16\pi}C^\alpha C_\alpha + L_s \biggr) \, h \; d^4 x  \label{e10} \end{equation}  where $L_s$ is the appropriate source Lagrangian density function.
Variation of (\ref{e10}) with respect to the tetrad results in
$$\int \Biggl[\frac{1}{16\pi}\biggl(2C_{(\mu ;\nu)}- 2C_\alpha \Upsilon^\alpha_{\; (\mu\nu)} - g_{\mu\nu}C^\alpha C_\alpha - 2g_{\mu\nu}C^\alpha_{\; ;\alpha} \biggr) - (T_{\rm s})_{\mu\nu}\Biggr] h\, h^{i\nu} \delta h_i^{\;\mu} \, d^4 x = 0$$
The $(T_{\rm s})_{\mu\nu}$ term is the usual source stress-energy tensor according to  standard procedures [11].  Thus
\begin{equation} C_{(\mu ;\nu)}- C_\alpha \Upsilon^\alpha_{\; (\mu\nu)} -\frac12 g_{\mu\nu}C^\alpha C_\alpha - g_{\mu\nu}C^\alpha_{\; ;\alpha} = 8\pi (T_{\rm s})_{\mu\nu} \label{e11} \end{equation}
and
\begin{equation} G_{\mu\nu}=  \biggl( \Upsilon^{\;\;\alpha}_{(\mu \;\; \nu ) ;\alpha}+\Upsilon^\alpha_{\;\;\sigma
 (\nu} \Upsilon^\sigma_{\;\; \mu ) \alpha} + \frac12 g_{\mu\nu}\Upsilon^{\alpha \beta
\sigma} \Upsilon_{\alpha \sigma \beta} \biggr) + 8\pi (T_{\rm s})_{\mu\nu}  \quad   \label{e12} \end{equation}
or
\begin{equation} \mathbf{G} \, =  \, 8\pi\bigl(\, \mathbf{T_f}\, + \, \mathbf{T_s} \, \bigr)  \qquad .  \label{e13} \end{equation}
The first term represents the free-field part of the total stress energy tensor which may be interpreted as the dark matter and/or dark energy part.  The second term due to the sources corresponds to ordinary matter and/or forces, such as the perfect fluid model.

\par \phantom{D} \par

\section{Spherically symmetric solutions.}

In this section we develop the formulae for the curvature vectors, metric, total stress-energy tensor and source stress-energy tensor.  These formulae will be utilized to find these quantities in our static universe model.  In spherical coordinates, an arbitrary spherically symmetric tetrad may be expressed by
\begin{equation}h^i_{\;\; \mu} =  \left[ \begin{array}{cccc}
\; e^{\Phi(r)} & 0 & 0 & 0 \\
0 &\; e^{\Lambda(r)}\sin \theta \cos \phi \; & \; r\cos\theta\cos\phi \; & \; -r\sin\theta\sin\phi \; \\
0 & e^{\Lambda(r)}\sin\theta\sin\phi & r\cos\theta\sin\phi    & \; \;r\sin\theta\cos\phi \\
0 & e^{\Lambda(r)}\cos\theta\qquad & -r\sin\theta\qquad & 0
\end{array}
\right]
\label{e31} \end{equation}
where the upper index refers to the row. The curvature vector for this tetrad field is given by
\begin{equation} C_\mu = \frac{e^{\Lambda}}r \biggl[ \; 0,  \; 2 - e^{-\Lambda}\bigl(r\Phi^\prime + 2\bigr) , \; 0 , \;0  \biggr]  \label{e32} \end{equation}
where components are in the order $[t,r,\theta,\phi]$ and the prime denotes the derivative with respect to $r$. The tetrad (\ref{e31}) leads to the metric
\begin{equation}ds^2= -e^{2\Phi(r)} dt^2 +  e^{2\Lambda(r)}dr^2+r^2d\theta^2 +r^2\sin^2\theta d\phi^2 \quad . \label{e33} \end{equation}
When $(r\Phi^\prime + 2)=2e^{\Lambda}$, then $C_\mu$ in equation (\ref{e32}) is identically zero and hence the field equations for the free field (\ref{fieldeq}) are satisfied. For the present discussion, we need a non-zero $\mathbf{T_s}$.

  The metric (\ref{e33}) leads to a diagonal Einstein tensor with nonzero elements:
\begin{equation}G^t_{\, t}\equiv -8\pi \rho  \; = \, \frac1{r^2}\bigl(-2re^{-2\Lambda}\Lambda^\prime + e^{-2\Lambda} - 1 \bigr) \;  = -\frac2{r^2}\frac{d}{dr}\biggl[\frac12 r \Bigl(1  - e^{-2\Lambda}\Bigr) \biggr]  \label{e34} \end{equation}
\begin{equation}G^r_{\, r} \equiv \; 8\pi p_r \; = \; \frac1{r^2}\Bigl( 2re^{-2\Lambda}\Phi^\prime + e^{-2\Lambda}-1\Bigl)  \label{e35} \end{equation}
\begin{equation} G^\theta_{\, \theta}\, =\; G^\phi_{\, \phi} \, \equiv \; 8\pi p_T \, = \; \frac{e^{-2\Lambda}}{r}\biggl(r\Phi^{\prime\prime}+r(\Phi^\prime)^2 - r\Phi^\prime \Lambda^\prime + \Phi^\prime - \Lambda^\prime \biggr)  \label{e36} \end{equation}

For our present discussion, we want to write the total stress-energy tensor (\ref{e13}) as a sum of a dark matter/energy stress-energy tensor and a source stress-energy tensor.
We will use the general spherical tetrad and the field equations which are derived from the Lagrangian (\ref{e10}) with $L_s=\rho_s(r)$, where $\rho_s(r)$ is the density as a function of $r$.  This Lagrangian with appropriate thermodynamic conditions leads to the usual perfect fluid stress-energy tensor \cite{dFC90}.  With a tetrad that corresponds to a co-moving frame ($h^0_{\; \mu} = 0$ for $\mu = \, 1, 2 \, , \, 3$ ), one finds \cite{Green11}
\begin{equation} \bigl(T_{\rm s}\bigr)^\mu_{\; \nu} = \left[
\begin{array}{cccc}
 -\rho_s \; & \; 0\; & \;0\; &\; 0\; \\
 0 & p_s & 0 & 0 \\
 0 & 0 & p_s & 0 \\
 0 & 0 & 0 & p_s
 \end{array}
 \right] \quad . \label{e40} \end{equation}
Using (\ref{e31}), we require that the radial and tangential pressures of the corresponding source stress-energy tensor (\ref{e11}) be equal.  This leads to the following differential equation (primes denote derivatives with respect to $r$):
\begin{equation} r^2\Phi^{\prime\prime} -\bigr(r^2\Lambda^\prime + r e^{\Lambda} \bigr)\Phi^\prime = 2 - 2 e^{2\Lambda} + 2r\Lambda^\prime  \label{e41} \end{equation}
After multiplying by an integrating factor, integration yields
\begin{equation}   \bigl(r\Phi^\prime + 2 \bigr)e^{-\Lambda} = 2 - \kappa re^{\int (r^{-1}e^{\Lambda}) }   \label{e42} \end{equation}
where $\kappa$ is arbitrary. This implies that
\begin{equation} C_\mu \, = \, \Bigl[ \; 0 \, , \, \kappa e^\Lambda e^{\int (r^{-1}e^\Lambda } \, , \; 0 \, , \; 0 \; \Bigr] \label{Cmu2} \end{equation}
After dropping a pure divergence term from the field equations (a standard procedure in field theory), we find that
\begin{equation}   8\pi \rho_s = \frac12 \kappa^2 e^{\;\int 2e^\Lambda/r}  \label{rho_s} \end{equation}
\begin{equation} 8\pi p_s = \frac{\kappa}r  e^{\;\int e^\Lambda/r} - \frac12 \kappa^2 e^{\; \int 2 e^\Lambda/r}    \quad . \label{p_s} \end{equation}
We also find that \begin{equation} C^\mu C_\mu = \kappa^2 e^{\;\int 2e^\Lambda/r} \, = 16\pi \rho_s \quad . \label{CmuCmu2} \end{equation}

\par \phantom{D} \par

\subsection{Particle motion in the spherically symmetric case.}

In order that particle motion may be investigated in our cosmological models, we develop here the general formulae governing such motion in a spherically symmetric solution of our field equations.  The equations of motion for a particle are based on adding an appropriate term to the Lagrangian.  Following \cite{dFC90} we add a term
\begin{equation}  L_{p} = \rho_{p}(x)  = \mu \int \delta_\epsilon^4(x-\gamma(s))(-u^\mu u_\mu)^{\frac12} ds   \label{e67} \end{equation}
to the Lagrangian of (\ref{e10}). The $\delta_\epsilon^4$ function approximates the Dirac delta function with a space-like volume of $\epsilon$ and we require that, in the limit as $\epsilon\to 0$, it equals the usual Dirac delta function. The mass of the particle will be denoted by $\mu$.  The function $\gamma(s)$ gives the path of the particle and its velocity is $u^\alpha=\frac{dx^\alpha}{d\tau}$.  We will use the "dot" notation for the components of $u^\alpha$, i.e. $u^\alpha = \langle \dot{t},\dot{r},\dot{\theta},\dot{\phi}\rangle$.  The condition  $T^{\beta \alpha}_{\; \;\;\; ; \beta }=0$ leads to (see \cite{dFC90} and \cite{Green11})
\begin{equation}  \; \frac{\mu}{\sqrt{-u^\nu u_\nu}} \, u^\beta u^\alpha_{\; ;\beta} \, = \; \delta^\alpha_1 F_p   \label{e68} \end{equation}
with $F_p \equiv \epsilon \, e^{-2\Lambda} \bigl(-p_R^{\; \prime} + \frac2r \, \bigl(p_T-p_R\bigr)\,\bigr) $.  When $\alpha\neq 1$, (\ref{e68}) implies  $u^\beta u^\alpha_{\; ;\beta} = 0$ which is the usual geodesic equation.

For the spherically symmetric metric given in (\ref{e33}) we find the $\theta$ component of (\ref{e68}).  After multiplying by $ \frac{\sqrt{-u^\nu u_\nu}}{\mu} $  we obtain
\begin{equation}   \ddot{\theta} + \frac2r \, \dot{r}\, \dot{\theta} - (\sin\theta \cos\theta) \dot{\phi}^2 =0   \label{e69} \end{equation}  We note that $\theta\equiv \frac{\pi}2$ implies $\dot{\theta} = 0$ and hence (\ref{e69}) yields $\ddot{\theta} = 0$.  Symmetry considerations imply that we may safely assign this value of $\theta$ since particle motion may be assumed to take place in a plane through the origin.  Henceforth assume that $\theta\, \equiv \, \frac{\pi}2$.

The $\phi$ component of (\ref{e68}) using metric (\ref{e33}) implies that
\begin{equation}   \ddot{\phi} + \frac2r \, \dot{r} \, \dot{\phi} \, = 0  \label{phiddot} \end{equation}
and this leads to
\begin{equation}   \dot{\phi} \; = \; \frac{L}{r^2} \label{phidot} \end{equation}
where $L$ is a constant interpreted as the conserved angular momentum.
From (\ref{e68}) we also find that the $\, t\,$ component is
\begin{equation}  \ddot{t} \, + \, 2 \dot{\Phi} \, \dot{r} \, = \, 0  \label{tddot} \end{equation}
and thus
\begin{equation} \dot{t} \, = \, E \, e^{-2\Phi} \label{tdot} \end{equation}
where $\Phi(r)$ is the function appearing in $h^0_{\, 0} \, $ (see \ref{e31}).

For the spherically symmetric metric (\ref{e33}) with $\theta \equiv \frac{\pi}2$, the $r$ component of the particle motion is determined by
\begin{equation}  \ddot{r} \, + \, e^{2\Phi-2\Lambda}\Phi^\prime (r) \, \dot{t}^2 \, + \, \Lambda^\prime (r) \, \dot{r}^2 \, - \, r \, e^{-2\Lambda}  \dot{\phi}^2 \, = \; \frac1{\mu} F_p \label{rddot}
\end{equation}
Using (\ref{phidot}) and (\ref{tdot}) along with the normalization $u^\nu u_\nu = -1$, we find that
\begin{equation}  \ddot{r} \, + \, (\Lambda^\prime + \Phi^\prime ) \, \dot{r}^2 \, + \frac{e^{-2\Lambda}}{r^3}\biggl[ r^3\Phi^\prime + L^2(r\Phi^\prime - 1 ) \biggr] \, = \; \frac1{\mu} F_p   \label{rddot2} \end{equation}

\par \phantom{D} \par

\section{Static Model for the Universe.}

In the standard cosmological model, space is assumed to be homogeneous and isotropic \cite{Jack07}.  These requirements are imposed on ordinary matter and ordinary pressures, implying that the density of mass-energy of ordinary matter is constant.  We, on the other hand, require that the overall, total mass-energy content be approximately constant.  Our foundational assumption is that
\begin{equation}   8\pi \rho \; \equiv \; 3\alpha^2 \label{rho_total} \end{equation}
where $8\pi$ is a convenient factor and $\alpha^2$ is a constant to be determined.
We also assume throughout the present discussion that the Milky Way Galaxy is close to $r=0$ (the calculations below are fairly accurate as long as the Milky Way Galaxy's $r$-value is less than $\frac13$ the radius of the universe.

From (\ref{e34}) and (\ref{rho_total}) we find that $\frac2{r^2}\frac{d}{dr}\Bigl[\frac12 r (1  - e^{-2\Lambda})\Bigr]=3\alpha^2$ thus \linebreak $\frac12 r(1-e^{-2\Lambda} ) = \frac12 \alpha^2 r^3 $
and so $1-e^{-2\Lambda} = \alpha^2 r^2$ and hence
\begin{equation} e^{2\Lambda} \; = \;  \frac{1}{1-\alpha^2r^2} \label{e2Lambda} \end{equation}
The integral in (\ref{e42}) can be easily evaluated resulting in $ \, e^{\int(\frac{e^\Lambda}r)} =  \Bigl( \frac{1-\sqrt{1-\alpha^2r^2}}{r} \Bigr)$. In order to match the weak field approximation, where the weak field is near $r=0$, we will choose $\kappa = 4$.  This implies that the curvature vector is nonzero which corresponds to saying that there must be a source with nonzero mass.  Using $\, \kappa = 4\,$, we find $ \kappa \, e^{\int(\frac{e^\Lambda}r)} =  4\Bigl( \frac{1-\sqrt{1-\alpha^2r^2}}{r} \Bigr) \, = \frac{4  \alpha^2 r}{1+\sqrt{1-\alpha^2r^2}} $.  Using (\ref{e42}), this leads to $r\Phi^\prime = -2 + (1-\alpha^2r^2)^{-\frac12} \Bigl[ 2\, -\, 4(1-\sqrt{1-\alpha^2r^2} \, ) \, \Bigr] \, = \, 2 \,- \, (1-\alpha^2r^2)^{-\frac12}$.  After integrating and choosing the constant of integration so that $g_{00} = -1$ at $r=0$ results in
\begin{equation}  e^{2\Phi} = \frac{\,(\,1+\sqrt{1-\alpha^2r^2}\,)^4}{16} \label{e2Phi}  \end{equation}
The resulting metric (in line element form) is given by
\begin{equation}  ds^2 \, = \; \frac{- \,(1+\sqrt{1-\alpha^2r^2})^4\, dt^2 }{16} + \frac{dr^2}{\, 1-\alpha^2r^2 \,} + r^2 d\theta^2 + r^2\sin^2\theta \, d\phi^2 \label{metric} \end{equation}
and the curvature vector (only nonzero component: $\;C_1 = \frac{4\alpha^2 r}{\, (1+\sqrt{1-\alpha^2r^2})\sqrt{1-\alpha^2r^2}\; }$) when contracted with itself yields
\begin{equation} C^\mu C_\mu \; = \; \frac{16 \alpha^4 r^2}{\;(1+\sqrt{1-\alpha^2r^2})^2 \,}  \end{equation}
From  (\ref{e2Phi}) and (\ref{rho_s}) we find that the density of the source is
\begin{equation}  8\pi \rho_s \; = \;  \frac{ 8 \alpha^4r^2}{(1+\sqrt{1-\alpha^2r^2})^2} \label{rho_s1} \end{equation}
and from (\ref{p_s}) we find that the pressure due to the source is
\begin{equation} 8\pi p_s \; = \; \frac{4 \alpha^2 (2\sqrt{1-\alpha^2 r^2} - 1)}{1+\sqrt{1-\alpha^2 r^2}  }  \label{p_s1} \end{equation}
For $r$ in the interval $[0,\frac1\alpha ]\,$, $\, p_s$ is positive only if $\, r \, > \, \frac{\sqrt{3}}{2\alpha} \,$.
In summary we find that (see (\ref{e40}) the source stress-energy tensor is given by: $\; 8\pi T_{{\rm s}\; \nu}^{\;\mu} \; = $
\begin{equation}
\left[\begin{array}{cccc}
 -\frac{ 8 \alpha^4r^2}{(1+\sqrt{1-\alpha^2r^2})^2}  & \; 0\; & \;0\; &\; 0\; \\
 0 & \frac{4 \alpha^2 (2\sqrt{1-\alpha^2 r^2} - 1)}{1+\sqrt{1-\alpha^2 r^2}  }   & 0 & 0 \\
 0 & 0 & \frac{4 \alpha^2 (2\sqrt{1-\alpha^2 r^2} - 1)}{1+\sqrt{1-\alpha^2 r^2}  }  & 0 \\
 0 & 0 & 0 & \frac{4 \alpha^2 (2\sqrt{1-\alpha^2 r^2} - 1)}{1+\sqrt{1-\alpha^2 r^2}  }
 \end{array}
 \right]  \label{Tsource} \end{equation}

The total radial pressure is easily found from (\ref{e35}) to be
\begin{equation}   8\pi p_r \; = \; \frac{-\alpha^2 (\, 5\sqrt{1-\alpha^2 r^2} + 1 \,)}{1+\sqrt{1-\alpha^2r^2}}  \label{p_r1} \end{equation}
and we see that $p_r$ is negative for $\, 0 \leq r \leq \frac1\alpha \;$.  We find from (\ref{e36}) that the tangential pressure is
\begin{equation} 8\pi p_T \; = \; \frac{- \alpha^2 (\, 7\sqrt{1-\alpha^2 r^2} - 1 \,)}{1+\sqrt{1-\alpha^2 r^2}}  \label{p_T1} \end{equation}
and we see that $p_T$ is only positive on $[0,\frac1\alpha ]$ when $\, r \, > \, \frac{4\sqrt{3}}{7\alpha}\,$.
The total stress-energy tensor ($\mathbf{T} = \mathbf{T_f} + \mathbf{T_s}$) is given by $\; 8 \pi T^\mu_{\; \nu} \; =$
\begin{equation} \left[
\begin{array}{cccc}
 -3\alpha^2  & \; 0\; & \;0\; &\; 0\; \\
 0 & \frac{-\alpha^2 (\, 5\sqrt{1-\alpha^2 r^2} + 1 \,)}{1+\sqrt{1-\alpha^2r^2}} & 0 & 0 \\
 0 & 0 & \frac{- \alpha^2 (\, 7\sqrt{1-\alpha^2 r^2} - 1 \,)}{1+\sqrt{1-\alpha^2 r^2}} & 0 \\
 0 & 0 & 0 & \frac{- \alpha^2 (\, 7\sqrt{1-\alpha^2 r^2} - 1 \,)}{1+\sqrt{1-\alpha^2 r^2}}
 \end{array}
 \right]  \label{Ttot}\end{equation}

The total density minus the density of the source may be interpreted as the density of the dark matter represented by $\rho_d$.  We find that
\begin{equation}  8\pi \rho_d \, = \, \frac{ \alpha^2 (\, 11\sqrt{1-\alpha^2 r^2} - 5 \, )}{1+\sqrt{1-\alpha^2 r^2}} \label{rho_d} \end{equation}
We may similarly solve for the radial pressure and tangential pressure of the field (which corresponds to dark energy).  The resulting free-field (dark matter/dark energy) stress-energy tensor is given by (note $\alpha^2$ is factored out): $ \; \; 8\pi T_{d \; \nu}^{\; \mu} \; = $
\begin{equation} \alpha^2 \left[
\begin{array}{cccc}
 \frac{ - 11\sqrt{1-\alpha^2 r^2} + 5 \, }{1+\sqrt{1-\alpha^2 r^2}}   & \; 0\; & \;0\; &\; 0\; \\
 0 & \frac{-\, 13\sqrt{1-\alpha^2 r^2} + 3 \,}{1+\sqrt{1-\alpha^2r^2}} & 0 & 0 \\
 0 & 0 & \frac{- 15\sqrt{1-\alpha^2 r^2} +5 \,}{1+\sqrt{1-\alpha^2 r^2}} & 0 \\
 0 & 0 & 0 & \frac{- 15\sqrt{1-\alpha^2 r^2} + 5 \,}{1+\sqrt{1-\alpha^2 r^2}}
 \end{array}
 \right] \label{Tdark} \end{equation}
We consider negative densities to be nonphysical, however in the world space, the forces and energy are determined by the total stress-energy tensor.  If  $11\sqrt{1-\alpha^2 r^2} \geq 5$ then $\, 1-\alpha^2 r^2 \geq \, \frac{25}{121}\,$ and thus $\alpha^2 r^2 \leq \, \frac{96}{121}\,$.  Thus we see that $\rho_d$ is positive only when $\, 0\; \leq \; r \;  \leq \; \frac{4\sqrt{6}}{11\alpha} \; \approx \; 0.8907 \times \frac1\alpha \;$.

It is an easy matter to calculate the  scalar curvature from (\ref{rho_total}), (\ref{p_r1}) and (\ref{p_T1}).  The result is
\begin{equation}  R \; = \; \frac{2\alpha^2(1+11\sqrt{1-\alpha^2r^2})}{1+\sqrt{1-\alpha^2r^2}}    \label{R1} \end{equation}
From (4) we see that the value of $C^\alpha C_\alpha$ and $R$ are related by an identity.  We thus look at our results from the point of view of general relativity (GR).  From the Lagrangian density for GR with a cosmological constant, $\Lambda_0$, we find that
\begin{equation} \frac1{16\pi} \Bigl( R - 2\Lambda_0) = \frac1{16\pi} C^\mu C_\mu + \rho_s \label{cc1} \end{equation}
where $\rho_s$ is given by (\ref{rho_s1}). Substituting our values for $C^\mu C_\mu$, $R$ and $\rho_s$ and solving for $\Lambda_0$ yields
\begin{equation}  \Lambda_0 \; = \; \frac{\, 3\alpha^2 (9\sqrt{1-\alpha^2 r^2}-5)\,}{1+\sqrt{1-\alpha^2 r^2}} \label{cc2} \end{equation}
Although not a constant, the values of $\Lambda_0$ are roughly constant for $r < \frac1{3\alpha}$ (see {\it Graphs} subsection below).

We also calculate the proper radius of our spherically symmetric universe.  Since we have a static model and when $dt = d\theta = d\phi = 0$ we have $,ds^2 = \frac{1}{1-\alpha^2 r^2} \, dr^2\, $, then
\begin{equation} \Delta s = \int_0^{\frac1\alpha} \frac1{\sqrt{1-\alpha^2 r^2}} \, dr = \frac{\pi}{2\alpha}  \label{Delta_s} \end{equation}
The coordinate time required for light to travel from the center to the edge is determined by the null ray condition, $ds^2 = 0$. If we also have $d\theta = d\phi = 0$ then solving (\ref{metric}) for $dt$  yields $dt \, = \, \frac{4\, dr }{\sqrt{1-\alpha^2 r^2}\, (\, 1+\sqrt{1-\alpha^2 r^2}\,)^2}$.  Integrating we find the coordinate time required from the edge of the universe to the center is
\begin{equation}  \Delta t \, = \, \int_0^{\frac1\alpha} \frac{4 \, dr }{\sqrt{1-\alpha^2 r^2} \, ( \, 1+\sqrt{1-\alpha^2 r^2}\,)^2 \,} \, = \, \frac{8}{3\alpha} \label{Delta_t}  \end{equation}

\subsection{Model values for the mass of the universe, fraction of ordinary matter, cosmological constant and redshifts.}

The total mass-energy of the universe is found from (\ref{rho_total}) to be
\begin{equation} M \, =  \, \int_{r\leq \frac1\alpha }\int\int \frac{3\alpha^2}{8\pi} \, dV \; = \; \frac32 \int_0^{\frac1\alpha } \alpha^2 r^2 dr \; = \; \frac{1}{2\alpha}   \label{M_total} \end{equation}
The total mass-energy of the source is found from (\ref{rho_s1}) to be
\begin{equation} M_s = 4 \int_0^{\frac1\alpha } \frac{\alpha^4 r^4}{(1+\sqrt{1-\alpha^2r^2})^2} \, dr \, = \; \frac{20-6\pi}{3\alpha} \; \approx \; \frac{0.3835}{\alpha}  \label{M_s} \end{equation}
We find that the ratio of these masses is given by
\begin{equation} \frac{M_s}{M} \;  \approx \, 0.7670  \label{Mratio} \end{equation}
If we find the fraction of mass-energy for a region near the center of the universe we get much smaller values.  For example on the region  where  $\; 0 \leq r \leq \frac{0.8}{\alpha}$,  we get $\frac{M_s}{M} \approx 0.086$.  Thus the percentage of ordinary matter is highly sensitive to the $r$ interval because of the high density of ordinary matter near $\frac1{\alpha}$ (see Figure \ref{fig:1} below).

An approximate value of radius of the entire universe in Planck units is $R_u\approx 2.7 \times 10^{61}\, l_p$.  We will use this value to determine the value of $\alpha$ and hence the values of the mass, densities, pressure, curvature and cosmological constant.  We caution that if and when a better value of $R_u$ is determined by observation then all of the results of this section should be updated.  We also assume due to (\ref{Delta_s}) and (\ref{Delta_t}) that the value of $r$  at the edge is actually half the currently reported value of $R_u$ (this assumption may not be true when a better values of $R_u$ is found).  We also assume that the Milky Way is near the center with an $r$-coordinate less than $\frac1{6\alpha}$.  Hence our starting point for our calculations is that
$\, \frac12 \cdot R_u = \frac1{\alpha}$
and we find
\begin{equation} \alpha \approx 7.41 \times 10^{-62}\, l_p^{\; -1}\end{equation}
An approximate  value for the total mass of the universe (\ref{M_total}) in Planck units is
\begin{equation} M = \frac1{2\alpha} \approx 6.75 \times 10^{60} \, l_p \end{equation}
The resulting value of the total density is
\begin{equation}  8\pi \rho = 3 \alpha^2  \approx 1.65 \times 10^{-122} \, l_p^{\; -2} \label{rho_tot_approx} \end{equation}
and the resulting value of $\Lambda_0$ at $r=0$ is
\begin{equation}  \Lambda_0(0) \; = \; 6\alpha^2 \approx 3.29 \times 10^{-122}\, l_p^{\; -2} \label{ccvalue} \end{equation}
This is close to currently accepted value of the cosmological constant \cite{BarrowShaw},\cite{Carroll}. Since our value depends heavily on our $r$-coordinate value corresponding to the edge of the universe which is equivalent to $\frac1{\alpha}$, the standard value may be easily obtained by a slight adjustment of this parameter.

The redshift for a photon emitted at some positive $r$ value and received near $r=0$ (where the metric is the flat space metric of special relativity) then the formula for the redshift $z$ is
\begin{equation}  z \equiv \frac{(\lambda_{rec} - \lambda_{em})}{\lambda_{em}} = |g_{tt}(r)|^{-\frac12} - 1  \label{reddef} \end{equation}
This leads to
\begin{equation} z \, = \; e^{-\Phi} - 1 \; = \; \frac{4}{(1+\sqrt{1-\alpha^2 r^2})^2} - 1 \; \, \geq \, 0  \label{z} \end{equation}
On the $r$ interval $\, [\,0,\frac1\alpha\, ]\,$ which corresponds to the entire universe, we find that
\begin{equation} 0 \; \leq \; z \; \leq \; 3   \label{red_interval}\end{equation}

\subsection{Graphs of Densities, Pressures, Curvature, Redshift and $\Lambda_0(r)$.}

Recall that we assume that the Milky Way Galaxy has an $r$-value close to zero.  Here we see evidence that as long as the Milky Way's $r$-value is less than about $\frac1{6\alpha}$ then our calculations and conclusions remain approximately correct.

\begin{figure}
\includegraphics[width=1.0\textwidth]{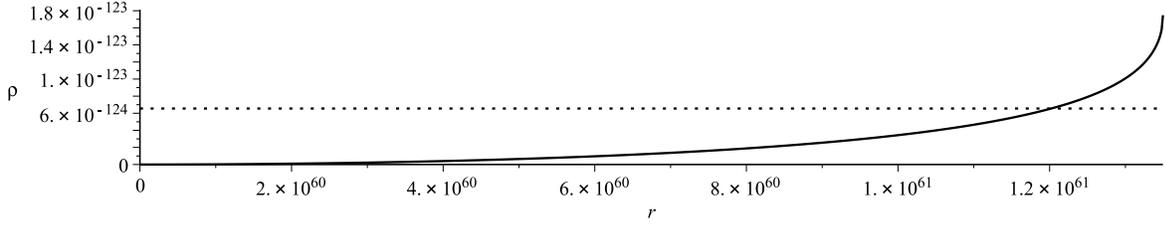}
\caption{Density of the Source (solid) and Total Density (dotted) with $\alpha =7.41 \times 10^{-62}\, l_p^{\;-1}$.  }
\label{fig:1}
\end{figure}

The graph of the density of the source for $\alpha = 7.41 \times 10^{-62}\, l_p^{\;-1}$  is shown in Figure \ref{fig:1}.  Note that the density of the source is very close to zero until $r$ is around $ 2.0 \times 10^{60}$ and this is the region (roughly) to which we assume the Milky Way galaxy belongs.  When $r$ is $1.0\times 10^{61}\, l_p$ or larger, the density increases more and more rapidly and as $r\to \frac1{\alpha} \approx 1.35 \times 10^{61} \, l_p$.  The value of the total density remains constant at a value of $\frac{3\alpha^2}{8\pi}$.

\begin{figure}
\includegraphics[width=1.0\textwidth]{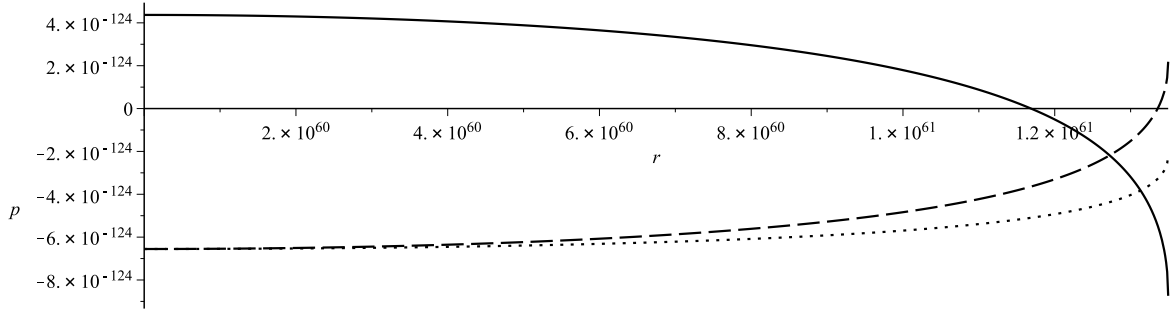}
\caption{Total Radial (dotted) and Tangential (dashed) Pressures and Pressure of the Source (solid) with $\alpha= 7.41 \times 10^{-62}\, l_p^{\;-1}$. }
\label{fig:2}
\end{figure}

The graphs of the total radial pressure and total tangential pressure are shown in Figure \ref{fig:2} along with the pressure of the source.  Both of the total pressures are negative and small on the entire interval from $r=0$ to $r=\frac1{\alpha}$. For $r$ smaller than about $2.0\times 10^{60}\, l_p$ their values are very close to $\frac{-3\alpha^2}{8\pi}\approx -6.55 \times 10^{-124}\, l_p^{\;-2}$ and then they gradually increase as $r$ approaches $\frac1{\alpha}$.  The pressure of the source is approximately $\frac{2\alpha^2}{8\pi}$ for $r<2.0\times 10^{60}\, l_p$.  As $r$ approaches $\frac1{\alpha}=1.35\times 10^{61}\, l_p$ the pressure of the source becomes negative and rapidly approaches $\frac{-4\alpha^2}{8\pi}\approx - 8.73 \times 10^{-124}\, l_p^{\;-2}$.

\begin{figure}
\includegraphics[width=1.0\textwidth]{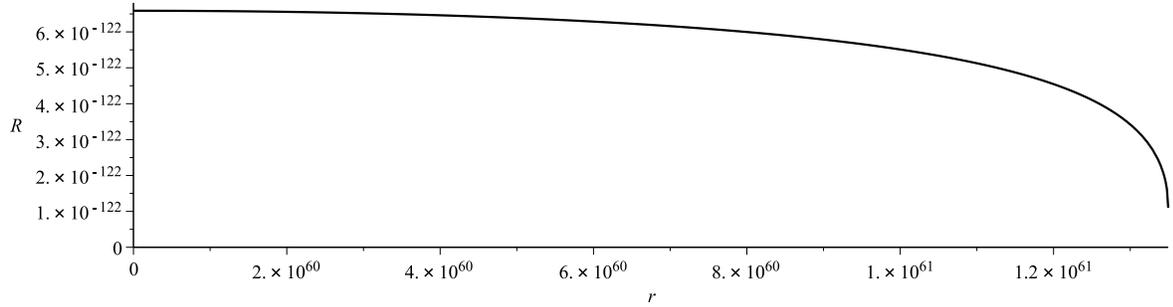}
\caption{Scalar Curvature with $\alpha = 7.41 \times 10^{-62}\, l_p^{\;-1}$.}
\label{fig:3}
\end{figure}

The graph of the scalar curvature which is shown if Figure \ref{fig:3} has a value of $12\alpha^2 \approx 6.58 \times 10^{-122}\, l_p^{\;-2}$ for $r$ less than $2.0 \times 10^{60}\, l_p$.  As $r$ approaches $\frac1{\alpha}$, the value of the scalar curvature decreases steadily and then rapidly, approaching a value of $R=   1.10 \times 10^{-122}\, l_p^{\;-2} $ at $r=\frac1{\alpha}$.

\begin{figure}
\includegraphics[width=1.0\textwidth]{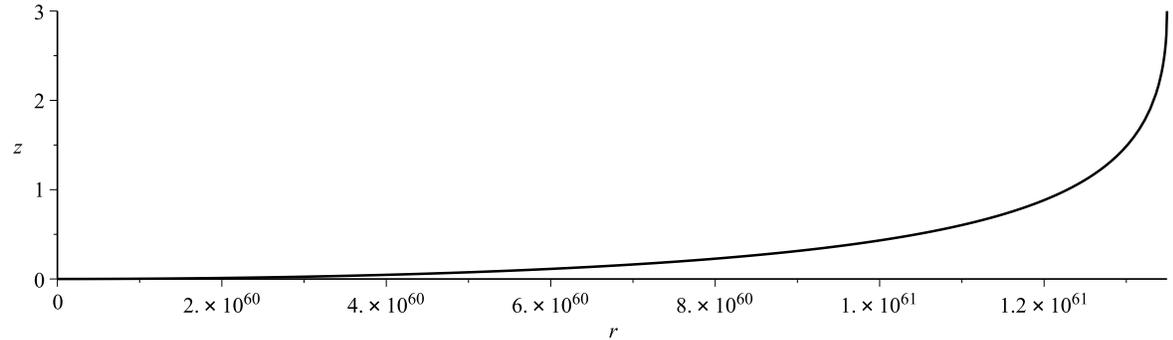}
\caption{Redshift $z$-values.}
\label{fig:4}
\end{figure}

The graph of the redshift, $z$, versus $r$ over the interval $0\, \leq \, r \, \leq \, \frac1{\alpha}$ is shown in Figure \ref{fig:4}. There is practically no redshift for $r$ less than $2.0\times 10^{60}\, l_p$.   We see a gradual increase, approximately linear increase in $z$ from $r=0$ to $r= 8.0 \times 10^{60}\, l_p$. For $r$  larger than this, the value of $z$ increases more and more rapidly.  The slope is actually infinite at $r=\frac1{\alpha}$.  As observed in redshift survey, the value increases with distance. The curve may be reasonably approximated by a quadratic.

\begin{figure}
\includegraphics[width=1.0\textwidth]{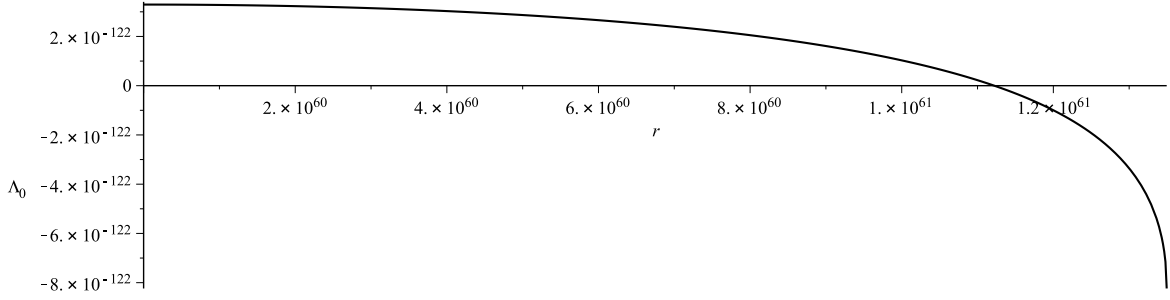}
\caption{Cosmological Constant ($\Lambda_0$) for $\alpha = 7.41 \times 10^{-62}\, l_p^{\;-1} $.}
\label{fig:5}
\end{figure}

Figure \ref{fig:5} shows the graph of $\Lambda_0 (r)$ as given by (\ref{ccvalue}) over the interval $0\, \leq r \, \leq \frac1{\alpha}$.  The value is close to $6\alpha^2 \approx 3.29 \times 10^{-122}\, l_p^{\;-2}$ for $r < 2.0 \times 10^{60}\, l_p$.  For $2.0 \times 10^{60}\, l_p < r < 1.12 \times 10^{61}\, l_p$, the value of $\Lambda_0$ drops steadily to zero. As $r$ approaches $\frac1{\alpha} \approx 1.35 \times 10^{61}\, l_p$ the value of $\Lambda_0$ drops rapidly to $-15\alpha^2 \approx  - 8.23 \times 10^{-122}\, l_p^{\;-2}$.

\subsection{Particle Motion in the Static Universe Model.}  From (\ref{e68}) using the functions $\Phi(r)$ and $\Lambda(r)$ for our static universe model (see (\ref{e2Lambda}) and (\ref{e2Phi}) ), we exhibit the equations of motion for a particle.  We will again restrict our attention to particles moving in a plane which contains the center of the universe and thus without further loss of generality we impose the condition $\theta \equiv \frac{\pi}2$.  The conserved angular momentum as always is given by (\ref{phidot}).  The conserved energy is related to $\dot{t}$ by (\ref{tdot}) which in this case gives
\begin{equation} \dot{t} \, = \, \frac{16 \, E }{(1+\sqrt{1-\alpha^2 r^2})^4} \label{tdot2} \end{equation}
The equation governing the $r$ coordinate in the spherically symmetric case is generally given by (\ref{rddot2}). Let the particle density, $\frac{\mu}{\epsilon}$ be denoted by $\hat{\rho}$.  In our static universe model, we get
\begin{eqnarray}  \ddot{r} \; + \;  \frac{\alpha^4 r^3\,\dot{r}^2}{(1-\alpha^2 r^2)(1+\sqrt{1-\alpha^2 r^2})^2}\; - \; \frac{\, L^2 \,  \sqrt{1-\alpha^2 r^2}\, (2-\sqrt{1-\alpha^2 r^2 })}{r^3} \quad \nonumber \\  \nonumber \\ \qquad   \;   - \; \frac{\, 2\alpha^2 r \sqrt{1-\alpha^2 r^2}\, }{1+\sqrt{1-\alpha^2 r^2}}   \; = \; \frac{- \, \alpha^6 r^3 \, \sqrt{1-\alpha^2 r^2} }{2\pi \, (1+\sqrt{1-\alpha^2 r^2})^3 \hat{\rho}}  \quad \label{rddot3} \end{eqnarray}

\subsection{ Radial acceleration on stationary particles.}
When $L=0$ and $\dot{r}=0$, then particle will move inward or outward depending on its density, $\hat{\rho}$.  When the particle density $\hat{\rho}$ matches the critical value of
\begin{equation} 8 \pi \, \hat{\rho} \, = \, 8 \pi \, \rho_{crit} \; \equiv \;  8 \pi \Bigl( \rho + p_r \Bigr) \; = \frac{2\alpha^4 r^2 }{(1+\sqrt{1-\alpha^2 r^2})^2} \label{rhocrit} \end{equation}
then we get $\ddot{r} = 0$ and thus the particle will not move. When the density is larger than $\rho_{crit}$ the term on the righthand side is small and thus the particle will be driven outward. For particles with densities smaller than $\rho_{crit}$, the term of the righthand side dominates and thus $\ddot{r}<0$ and so the particle is driven inward (blue shift D\"oppler effect).  Suppose in the $L=0$, $\dot{r}=0$ case that at a given value of $r=r_1$ that  $\hat{\rho} \, = \, \frac{\rho_{crit}}{\beta_1}$ for some $\beta_1>0$.  Then in this case,
\begin{equation} \ddot{r}\,\biggl|_{r=r_1} = \frac{\; 2 \,( \, 1-\beta_1 \,)\, \alpha^2 r \, \sqrt{1-\alpha^2 r^2}}{1+\sqrt{1-\alpha^2 r^2} }\,\biggl|_{r=r_1} \qquad  (\; {\rm if } \quad L = 0 \; {\rm and } \; \dot{r}= 0 \; )  \label{rddot4} \end{equation}
If $\, 0<\, \beta_1 <1 \,$ (i.e., $\hat{\rho}>\rho_{crit}\;$) then $\ddot{r}>0$ and there would be a redshift D\"oppler effect.  We would expect $\beta_1$ to be very small for most particles, however, the theory implies (see \cite{Green11}) that every massive object would have a halo and the particle and its halo are inseparable.  Thus the object's density is greatly reduced by the halo.  The halo has a stabilizing effect.

\subsection{ Circular Orbits.}
We see that circular orbits with a constant value of $r$ are possible because of the repulsive nature of the $L^2$ term (thus (\ref{rddot3}) is consistent with $\ddot{r} = 0 $ and $\dot{r}=0$). Now consider the case when $L\neq 0\;$ and $\dot{r}=0$, but $\hat{\rho}<\rho_{crit}$, where $\rho_{crit}$ is given by (\ref{rhocrit}).   We again will express $\hat{\rho}$ in terms of $\beta_1$ by $\hat{\rho}= \frac{\rho_{crit}}{\beta_1} \; $ with $\; \beta_1 > 1 \, $.  Then using the right-hand side of (\ref{rddot4}) and adding the nonzero $L$-term we would get
\begin{equation}  \ddot{r} \,\biggr|_{r=r_1} = \,  \frac{\;  2 \,( \, 1-\beta_1 \,)\, \alpha^2 r_1 \, \sqrt{1-\alpha^2 r_1^2}}{1+\sqrt{1-\alpha^2 r_1^2} } + \frac{\, L^2 \,  \sqrt{1-\alpha^2 r_1^2}\, (2-\sqrt{1-\alpha^2 r_1^2 })}{r_1^3} \label{rddot6}  \end{equation}
For a circular orbit at $r=r_1$, we find that
\begin{equation} L^2 \; = \; \frac{2\, (\, \beta_1 - 1 \, ) \, \alpha^2 r_1^4 }{(1+\sqrt{1-\alpha^2 r_1^2})(2-\sqrt{1-\alpha^2 r_1^2})} \label{Lsquared2} \end{equation}
with $\beta_1$ assumed to be larger than 1.

\subsection{ Analysis of Pure Radial Motion in the $\,(\, r \, , \, \dot{r} \, )\,$ Plane.  }  
\begin{figure}
\includegraphics[width=5cm,height=0.98\textwidth,angle=270]{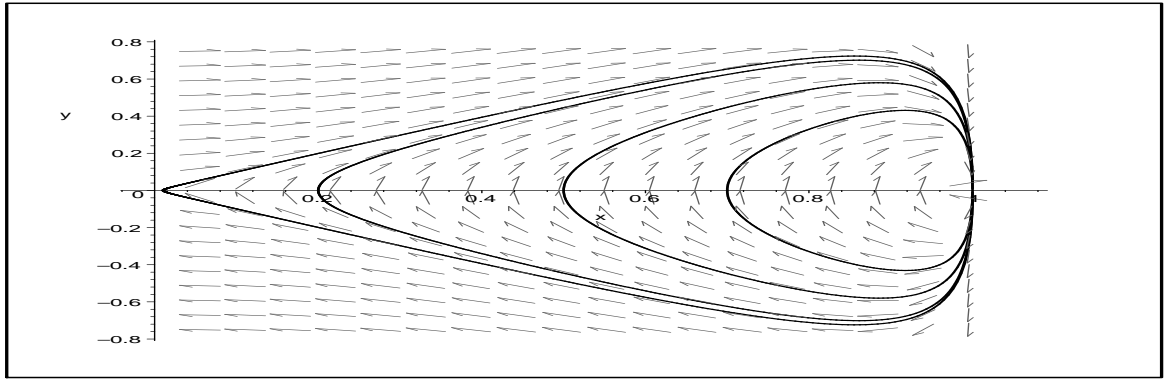}
\caption{Phase plane of $X$ vs. $Y$ on $[0,1] \times [-0.8,0.8]$ with $\beta=2$. The only fixed points are at $(0,0)$ and $(1,0)$.}
\label{fig:6}
\end{figure}

For this example we will assume the particle density is a constant given by $\, 8 \pi \hat{\rho} \, \equiv \, 2\, \beta^2 \alpha^2$, where $\alpha \approx 7.41 \times 10^{-62}\, l_p^{\;-1}$ and $\beta$ is a constant (different setup from (\ref{rddot4})).  Thus
\begin{eqnarray} \ddot{r} = &  \frac{-\alpha^4 r^3 \dot{r}^2 }{(1-\alpha^2 r^2)(1+\sqrt{1-\alpha^2 r^2})^2} + \frac{2\alpha^2 r \sqrt{1-\alpha^2 r^2}}{1+\sqrt{1-\alpha^2 r^2 }} - \frac{2\alpha^4 r^3 \sqrt{1-\alpha^2 r^2}}{\beta^2 (1+\sqrt{1-\alpha^2 r^2})^3}         \nonumber \\
& \qquad \label{rddot5} \end{eqnarray}
We attempt convert this to a system and scale the time variable as follows:  $X \equiv \alpha r$, $\tau \equiv \frac1\alpha \, t$ and $Y \equiv \frac{dX}{d\tau}$.  The resulting system is
\begin{eqnarray}  \frac{dX}{d\tau} \, &= \, Y \qquad \qquad \label{XYsys} \\
 & \nonumber \\
\frac{dY}{d\tau} \, & = \,  \frac{-X^3Y^2}{(1-X^2)(1+\sqrt{1-X^2})^2} \, + \frac{2X\sqrt{1-X^2}}{1+\sqrt{1-X^2}} - \frac{2X^3\sqrt{1-X^2}}{\beta^2 (1+\sqrt{1-X^2})^3} \nonumber \end{eqnarray}
Using linearization, the fixed point at $(X,Y)=  (0,0)$ is seen to be a saddle type with a stable and unstable eigenvector.  Interpretation:  the fixed point at $r=0$ is unstable and particles given a gentle push will move away.  The fixed point at $(1,0)$ corresponding to the edge of the universe is a singular point and linearization fails.  In the limit as $X\to 1^-$, we find that $\frac{dY}{d\tau}\leq 0$.  The phase-portrait agrees with this analysis (see Figure \ref{fig:6}  and Figure \ref{fig:7}).  These are the only fixed points when $\beta \geq 1$ but when $0\leq \beta < 1$, there is another fixed point at $(X,Y)=(\frac{2\beta}{1+\beta^2} \, , \,0 \, )$.  This is a center type at which the density given by $\hat{\rho}=\frac{\beta^2\alpha^2}{4\pi}$ is equal to $\rho_{crit}$ as given by (\ref{rhocrit}).  Interpretation:  for particles with small densities, the motion will be oscillatory around an $r$-value determined by (\ref{rhocrit}).  Since these densities are very small, we find that only For particles with larger densities $\beta>1$, the motion is toward $r=\frac1\alpha$ (or $X=1$).  Only a cataclysmic event would cause such a particle to return (along the bottom trajectory).   The interesting result of the phase-portrait is that the velocity of typical particles accelerated from rest may easily rise to about $\frac{7}{10}$ of the speed of light or higher.

\subsection{ Redshifts of Particles.}
The usual D\"oppler effect is  due to a difference in the velocity alone and its value as observed by a stationary observer in close proximity will not be affected by the gravitational shift.  Let $v$ represent the velocity of the particle divided by the speed of light in the local Lorentz frame.  We have the well-known formula $z_{Dop}= \sqrt{\frac{1+v}{1-v}}-1$.  As the particle's light is transmitted to the Milky Way (or other observer near the center), it is further redshifted.  Let $\lambda_e$ represent the wavelength of the emitter, $\lambda_{e^\prime}$ the wavelength measured nearby by a stationary observer (relative to center) and let $\lambda_o$ represent the observed wavelength by the final observer.  Then
\begin{eqnarray} z_{tot} &=  \frac{\lambda_o}{\lambda_e}-1 = \frac{\lambda_o}{\lambda_{e^\prime}}\cdot \frac{\lambda_{e^\prime}}{\lambda_e} - 1  = \frac{\lambda_o}{\lambda_{e^\prime}}\biggl(z_{Dop}+1\biggr) - 1 \nonumber \\
 & = \biggl(z_{grav} + 1 \biggr)\sqrt{\frac{1+v}{1-v}}  \; - \;  1  \label{z_tot}\end{eqnarray}
where $z_{grav}$ is the gravitational redshift of (\ref{z}).

\begin{figure}
\includegraphics[width=5cm,height=0.98\textwidth,angle=270]{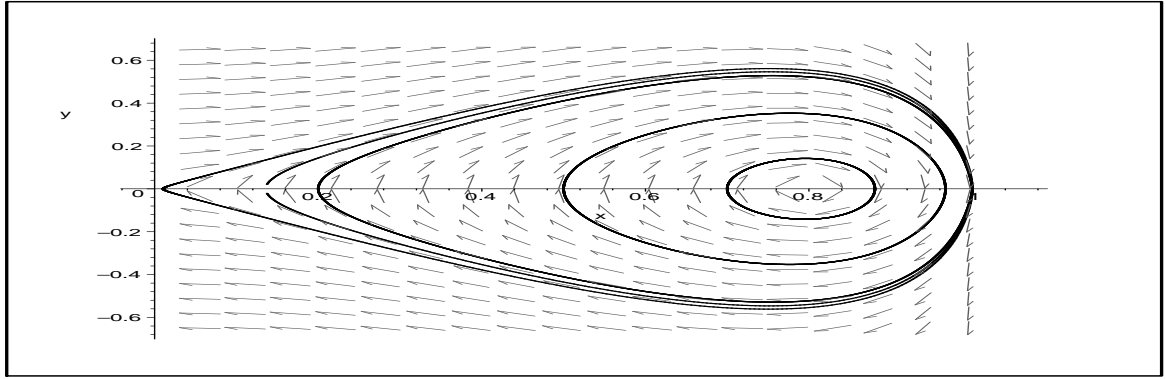}
\caption{Phase plane of $X$ vs. $Y$ on $[0,1] \times [-0.7,0.7]$ with $\beta=0.5$. The only fixed points are at $(0,0)$, $(1,0)$ and the center at $(0.8,0)$. }
\label{fig:7}
\end{figure}

The $\beta<1$ case, although interesting, does not correspond to realistic densities. Only large values of $\beta$ correspond to realistic densities.  From the dynamics we see that there is a correlation between the maximum radial velocity and the $r$-value at which it occurs (see Figure \ref{fig:7}).   In Table 1 we use an estimated value of $v$ along with the $z_{grav}$ given by (\ref{z}) at a given value of $X=\frac{r}{\alpha}$ to calculate $z_{tot}$ as given by (\ref{z_tot}).

The results of Table 1 do not include exceptional cosmic events.  For an exceptional event at $X=0.98$ occurred with an outward velocity of $v=0.9$, then $z_{tot} = 12.13$.  Near $r=0$, the same exceptional event would create a value of $z_{tot} = 3.36$.  Thus we see that large redshifts are likely for $r$ near the edge of the universe, i.e., at extremely large distances from earth.

\begin{table}
\caption{  Particle $z$-Values for $\beta>1$    }
\label{tab:1}
\begin{tabular}{llll}
\hline\noalign{\smallskip}
\qquad $X$ & $z_{grav}$ & Est. max. value of $v$  & $ \; z_{tot}  $ \\
\noalign{\smallskip}\hline\noalign{\smallskip}
\quad $0.70$ & 0.3613   &  \qquad 0.6   &  1.72  \\
\quad $0.80$ & 0.5625  & \qquad 0.7    &  2.72   \\
\quad $0.90$ & 0.9401  & \qquad 0.6    &  2.88   \\
\quad $0.96$ & 1.4414  & \qquad 0.4    &  2.73   \\
\quad $0.98$ & 1.7823   & \qquad 0.3    &  2.76  \\
\quad $0.999$ & 2.6650 & \qquad 0.05    &  3.85 \\
\noalign{\smallskip}\hline
\end{tabular}
\end{table}

\par \phantom{D} \par

\subsection{Cosmic Microwave Background.}
In this section, we give heuristic arguments that indicate that our theory in not inconsistent with observational data about the cosmic microwave background (CMB).
The temperature of the CMB is nearly uniform in all directions (isotropic) and is approximately $2.7^{\rm \, o}$K.  This low temperature is partly due to a redshift.

\subsubsection{Explanation of Temperature of the CMB.}
The results of the previous section, particularly (\ref{rddot4}), show that at the edge of the universe ($r=\frac1\alpha$), it is possible to have stationary particles indicated by $\dot{r}=0\,$ and $\,L=0\,$. At the same time, the particles at $r=\frac1\alpha$ are under negative radial pressure and this tends to hold a particle in place via tension. According to (\ref{Ttot}), at $\, r\, =\, \frac1\alpha \,$, $\, 8\pi\, p_r = - \alpha^2 $ and $\, 8 \pi \, p_T = \, \alpha^2$.  These pressures tend to prevent a stationary particle at $r=\frac1\alpha$ from moving.  The expected value of $\rho_{crit}$ corresponds to $\beta=1$ and from (\ref{rddot5}), we see that if $\dot{r}$ is nonzero when $r=\frac1\alpha$ then there is a extremely strong, inward acceleration.  This has the effect of removing (similar to evaporation) any particle that has a significant nonzero kinetic energy. Thus a low temperature around $2.7^{\rm\, o}$K is consistent with this model.

\subsubsection{Explanation of Anisotropies of the CMB.}
According to (\ref{rddot4}) when $\beta_1 < 1$, a stationary particle will move outward.  According to (\ref{rddot5}), if $\beta\geq 1$ (note: different from $\beta_1$), then the only stable fixed point is at $r=\frac1\alpha$, i.e., the edge of the universe.  Thus any object comprised of ordinary matter, such as a galaxy, would have densities large enough so that $\beta_1 < 1$ and also $\beta\geq 1$.  Why is it that all the ordinary matter is not driven to the edge of the universe?  The answer is that ordinary matter teams up with other ordinary matter to make bubbles.  The interior of the bubble is nearly empty space and these voids are a well-known feature of the universe.  By locking together in a bubble, the overall density of the bubble is then sufficiently low that, on average, $\beta_1 = 1$ for the bubble.

The bubbles are nearly spherical near $r=0$ and as $r$ approaches $\frac1\alpha$, the bubble would become flattened on the outer side, similar to a raindrop.  We will approximate the bubbles as spherical balls of radius $\hat{r}_n$ centered at $r_n$, where the index $n$ is zero at the center where $r=0$ and $r_{n+1} > r_n$ for $n=0,1,2,\dots$.  We will also make the approximation that the density of the surface of a bubble is a constant, represented by $k$.  The bubble at the center has a radius $\hat{r}_0$ and is centered at $r_0=0$.  Using the formula for the critical density (\ref{rhocrit}), which corresponds to the inertial mass, we find that when $\alpha r$ is small, the total inertial mass in the center bubble approximately $\frac16 \alpha^2 \hat{r}_0^3$ (assumes $\alpha \hat{r}_0$ is small).  Thus $4\pi \hat{r}_0^2 k = \frac16 \alpha^2 \hat{r}_0^3$ from which we find that $k= \frac{\alpha^2 \hat{r}_0}{24\pi}$.  For the remaining bubbles, we approximate the mean critical density of the bubble using the value of $\rho_{crit}$ at the center $r_n> \, \hat{r}_0$, which implies that $4\pi \hat{r}_n^2 k = \frac{4\pi\hat{r}_n^3  }3 \times \frac{\alpha^4 r_n^2}{(1+\sqrt{1-\alpha^2 r_n^2})^2}$.  We thus find that
\begin{equation} \alpha \hat{r}_n \; = \; \frac{(1+\sqrt{1-\alpha^2 r_n^2})^2}{\alpha^2 r_n^2} \times \frac{3k}{\alpha} \; =
\; \frac{(1+\sqrt{1-\alpha^2 r_n^2})^2}{\alpha^2 r_n^2} \times \frac{\alpha \hat{r}_0}{8\pi}  \label{bubbleradius}\end{equation}
We estimate that $\alpha \hat{r}_0 \approx 0.006$ (about 80 Mpc), which leads to a maximum void diameter of 400 Mpc.  Bubbles near the edge of the universe would have a radius of approximately $\frac1{30}$ of $\hat{r}_0$ and thus $\alpha \hat{r}_n \approx 0.0002$.  At the center of the universe, the measure of angle subtended by this last void would be approximately 0.0004 radians or $0.023^{\rm o}$. We propose that these voids are the source of the anisotropies of the CMB spectrum.

\par \phantom{D} \par

\section{Conclusion.}

For the theory based on the conservation group, we have developed a static model of the universe.  From our fundamental assumption that the total mass-energy is constant, we use the weak-field approximation to determine a spherically symmetric solution of our field equations that shows promise as a model of the universe.  We also note that this solution is stable, with a wide range of parameter values ($\alpha$ values) that lead to realistic models.  We find a universe model that is nearly isotropic when $r< \frac1{3\alpha}$.  The density of ordinary matter increases with $r$ to its greatest value at $\frac1{\alpha}$. Remarkably, we find that the usual gravitational redshift formula leads to redshift values that steadily increase with $r$ approximately quadratically and also are in the range of $z$ values that are observed.  This, so called, acceleration may be thus explained by this model.  We also have a plausible explanation of the value of the cosmological constant.  Furthermore, the theory may establish Mach's Principle. Particle equations of motion indicate that circular orbits are possible.  Pure radial motion is seen to depend on the particle's density.   We have shown that redshifts for particle motions are favored and tend to increase with distance and the calculation of $z_{tot}$ illustrates that large redshifts are possible for exceptional events near the edge of the universe.  Our discussion also suggests that our model in not inconsistent with CMB observations, although a full analysis may show otherwise.  The necessity of bubbles or voids whose diameter decreases as $r$ approaches the edge value of $r=\frac1\alpha$ may explain the CMB anisotropies.  Our model solves many of the problems of cosmology without the big bang and without cosmic inflation.  If our theory is correct, there would be no horizon problem or flatness problem.

\subsection*{Acknowledgments}  The author thanks Peter Musgrave, Denis Pollney and Kayll Lake for the GRTensorII software package which was very helpful.  Also, thanks to Professor Mark Spraker for his comments and encouragement.  The author thanks the University of North Georgia  for travel support.

\end{document}